\documentclass[12pt]{iopart}
\usepackage{color}
\usepackage{graphicx}
\usepackage{graphicx,epsfig}
\usepackage{tabularx}
\newcommand{\Rmnum}[1]{\expandafter\@slowromancap\romannumeral #1@}

\def\cs2{c_{s}^{2}}

 \def\be   {\begin{equation}}   \def\ee   {\end{equation}}
 \def\ba   {\begin{array}}      \def\ea   {\end{array}}
 \def\bea  {\begin{eqnarray}}   \def\eea  {\end{eqnarray}}
 \def\bean {\begin{eqnarray*}}  \def\eean {\end{eqnarray*}}

\begin{document}

\title{Tilt and Running of Cosmological Observables in Generalized Single-Field Inflation}

\author{Nicola Bartolo$^{1,2}$, Matteo Fasiello$^{3}$, Sabino Matarrese$^{1,2}$ and 
Antonio Riotto$^{2,4}$}
\vspace{0.4cm}
\address{$^1$ Dipartimento di Fisica ``G. Galilei'', Universit\`{a} degli Studi di 
Padova,  via Marzolo 8, I-35131 Padova, Italy} 
\address{$^2$ INFN, Sezione di Padova, via Marzolo 8, I-35131 Padova, Italy}
\address{$^3$ Dipartimento di Fisica ``G. Occhialini'', Universit\`{a} degli Studi di Milano Bicocca and INFN, Sezione di Milano Bicocca, Piazza della Scienza 3, I-20126 Milano, Italy}
\address{$^4$ CERN, Theory Division, CH-1211 Geneva 23, Switzerland\\
\vskip 0.5cm
DFPD-2010-A-17\,\, CERN-PH-TH/2010-238}
\vskip 0.5cm
\eads{\mailto{nicola.bartolo@pd.infn.it}, \mailto{matteo.fasiello@mib.infn.it}, 
\mailto{sabino.matarrese@pd.infn.it} and \mailto{riotto@mail.cern.ch}}

\begin{abstract}
Employing an effective field theory approach to inflationary perturbations, we analyze in detail the effect of curvature-generated Lagrangian operators on various observables, focusing on their running with scales. At quadratic order, we solve the equation of motion at next-to-leading leading order in a generalized slow-roll approximation for a very general theory of single-field inflation. We derive the resulting power spectrum, its tilt and running. We then focus on the contribution to the primordial non-Gaussianity amplitude $f_{NL}$ sourced by a specific interaction term. We show that the running of $f_{NL}$ can be substantially larger than what dictated by the slow-roll parameters.

\end{abstract}
\newpage

\tableofcontents

\section{Introduction}
Inflation \cite{guth,toni1} stands as one of the main pillars of modern cosmology as it naturally solves the so-called flatness, horizon and monopoles problems.
It also explain the production of density perturbations in the early Universe which then lead to Large Scale Structures \cite{lss1}-\cite{lss5} in the distribution of galaxies and temperature anisotropies in the Cosmic Microwave Background (CMB) \cite{smoot92}-\cite{wmap5}. In recent years, many different inflationary models have been studied and the advent of new generation experiments (the launch of the Planck satellite~\cite{Pl,Mand} and the continued analysis of WMAP data~\cite{kom}) pushes further the research in this direction. Indeed, one can now hope that with improved sensitivity to deviations from Gaussian statistics, we might be able to probe deeper into the dynamics of the inflaton~\cite{reviewk}. This translates into including and studying 3-rd and higher order (self) interactions in the inflationary Lagrangian \cite{bisp, bispectrum, trispectrum,loop}. A detailed analysis has been done for a huge variety of models (see \cite{chenrev} for a comprehensive and updated review) with precise predictions for the amplitude and the shapes of correlation functions. In fact, in performing these investigations, one is aiming at predictions for observables such as the scalar spectral index, its running, the primordial bispectrum amplitude, $f_{NL}$, and its running. The work we present here fits in this context and its goal is to highlight the contribution to these observables coming from 
specific operators of a very general effective field theory describing inflation driven by a single scalar degree of freedom.  A formalism that allows a unified approach to inflation, by means of effective field theory techniques,  has been presented in \cite{luty, eft08}. The prescription of \cite{eft08} is appealing for several reasons: one can see there is a clear cut dictionary between a given inflationary model and a specific linear combination of operators in the effective Lagrangian obtained by turning on and off some coefficients (later indicated with $M$'s) that regulate the weight of the operators. In fact, most of these coefficients turn out to be in principle free parameters (with the sole requirement for them to be smaller than the mass of the underlying theory) and this clearly enlarges the region of the parameters space than can be spanned by the theory. Specific inflationary models would, on the other hand, put strong bounds on several of these coefficients. 
Besides this fact, the effective Lagrangian naturally captures extrinsic curvature-generated operators that are sometimes neglected in the literature. These terms should definitely be studied as, in fact, their contribution can be relevant \cite{b,t} and therefore they can also also significantly increase the dimensions of the parameters space of the theory. These are the specific operators this paper will mainly focus on. One more advantage that comes with employing the proposed setup is of calculational nature: in the so called \textit{decoupling} regime (which implies to work in a specific energy range)  the dynamics of the metric decouples from the one of the scalar that drives inflation thus rendering the Lagrangian itself and the higher order correlators much easier to handle and calculate. This mechanism is very reminiscent of what happens in standard quantum field theory and goes under the name of equivalence theorem.\\
Interestingly, the effective approach proves itself useful already at quadratic order in the perturbations in that it automatically generates an equation of motion whose solution will encompass the classical wavefunction of many inflationary theories in the appropriate limits. We present here in detail such a solution. First, we give results at leading order in generalized slow-roll parameters for the power spectrum, its tilt $n_s$ and running. Secondly, we solve the equation of motion at next-to-leading order in slow-roll, then we specialize to the Ghost Inflation case and obtain tilt and running as well.\\
The amplitude $f_{NL}$ has already been calculated in a number of papers \cite{eft08,tilted,3pt,ssz05} within the effective field theory set up. Here in particular, we use the results of \cite{b} to calculate the running of the non-Gaussianity (NG). From (almost) scale invariance arguments one expects the latter to be a weak effect but it is nevertheless quite important. In fact, one expects the scale invariance to be eventually broken and also wants to use the running \cite{chen-run}-\cite{toni} as a tool to distinguish between inflationary models which generate the same predictions, even at the level of the bispectrum shape-function. \\ The  paper is organized as follows: in \textit{Section 2} we solve at leading order the generalized equation of motion, we compare it with known solutions in the proper limits and comment on the wavefunction behaviour deep inside the horizon and in the crossing-horizon region; in \textit{Section 3} we give the general quadratic solution up to next-to-leading order in slow-roll, we then specialize to Ghost Inflation and calculate the tilt and the running of the power spectrum; in \textit{Section 4} we calculate the running of the $f_{NL}$ portion generated by an intriguing third-order interaction term introduced in~\cite{b}, weighted by the $\bar M_6$ coefficient, and  show that it can be significant; in the \textit{Conclusions} we summarize our results and comment on further work; in the \textit{Appendix} we give explicit expressions for several quantities whose form has been kept compact in the main text for the sake of simplicity.\\

\section{Solution to the equation of motion for the scalar}
We here solve the equation of motion for the second-order effective Lagrangian derived in \cite{eft08} (see also \cite{b,ssz05}) at 
leading order in slow-roll:
\bea
\fl \mathcal{L}_2=a^3\Big(M_{P}^{2}\dot H (\partial_{\mu} \pi)^2 
+ 2 M_2^4{\dot\pi}^2 + \bar M_1^3 H \frac{(\partial_i \pi)^2 }{2 a^2}  
+ \frac{\bar M_2^2}{2}\frac{1}{a^4} (\partial_i^2\pi)^2+\frac{\bar M_3^2}{2} \frac{1}{a^4}(\partial_{ij}\pi)^2 \Big) .
\label{l2}
\eea

First though, a few comments on the above equation are in order.
\begin{itemize}
\item The procedure according to which the Lagrangian was obtained is outlined in \cite{eft08} and allows for a very general expression for inflation driven by a single scalar degree of freedom. In short, Eq.~(\ref{l2}) is the most general second-order Lagrangian in unitary (co-moving) gauge provided the approximate symmetry of the underlying theory is such that only derivative terms of $\pi$ appear in the action (see also \cite{w-e} for an in-depth analysis of these issues). To first order, (which is all we need here) the scalar $\pi$ is lineraly related to the dimensionless gauge invariant quantity $\zeta$ by $\zeta=-H \pi$.
\item In full generality the $M$ coefficients above should be time dependent (we will deal with such a case in \textit{Section 3}). However, if one is only interested in performing leading-order calculations, then, due to a generalized slow-roll approximation, one can safely consider them as constant.
\item Note that there is no trace of metric perturbations in Eq.~(\ref{l2}). This is because we are working in the so called \textit{decoupling regime}: for a sufficiently high energy range the dynamics of the scalar degree of freedom which drives inflation is decoupled from gravity. This is the so-called equivalence theorem at work. It suffices to say here that one can safely work with Eq.~(\ref{l2}) assuming $E > \epsilon^{1/2} H$, where as usual $\epsilon= \dot H /H^2$, and $ E > M_2^2/M_{Pl}$ .
\item Note for example that for $M_{2}=0=\bar M_{1,2,3}$ one re-obtains the usual quadratic Lagrangian with the speed of sound $c_s^2 =1$ and the standard oscillatory solution to the equations of motion. On the other hand, switching on the $M_{2}$ term amounts to allowing $c_s^2 <1$  models, $1/c_s^2 =1- 2M_2^4/(M_{Pl}^2 \dot H)$, which have been proven to give larger non-Gaussianity \cite{eft08,DBI,chen-bis,4pp,chen-tris}. Working in the de Sitter limit and  turning on $\bar M_{2,3}$ one rediscovers Ghost Inflation \cite{ghost}. This very same procedure enables one to capture all the corresponding models at higher orders as well. The list of correspondences can be continued with K-inflation \cite{mukh1, mukh2} theories and others proving that the effective action approach naturally provides a more unifying perspective on inflationary models.
\item The action in Eq.~(\ref{l2}) and, more importantly, its higher order counterparts, are generally written with large non-Gaussianities in mind. In the effective theory language, generalizing what has been done for DBI models, this corresponds to putting some constraints on the parameter space spanned by some time-independent (at leading order in slow-roll) coefficients in the action. A case in point is requiring a small speed of sound: this assumption, which can in general lead to to large NG, automatically translates into bounds on the values of the coefficients driving quadratic operators in the theory Lagrangian. A speed of sound different than unity necessarily generates a different weight, for example in Fourier space, between time-like and space-like derivatives acting on the scalar. A straightforward generalization of this argument shows that,  
 in writing  each $M, \bar M$ coefficient driving a specific operator in the Lagrangian, there is a meaningful criterion for a selection as to which are the leading terms to write down. To illustrate how one goes about determining which contributions fit in the action, we now consider a simple example. Let's take terms that are multiplied by  the $\bar M_1^3$ coefficient. In principle we should see in the above Lagrangian the contribution:
 \be
 - \bar M_1^3 ( -H (\partial_i \pi)^2 /a^2+ 6 H {\dot \pi}^2) .\label{mbarra1}
 \ee
On the other hand, we will show below that, in comparing terms of the same perturbative order within the same $M$ coefficient one just counts the number of space and time derivatives: the term with the highest number of space derivatives will be the dominating one. This is reminiscent of the fact that in DBI theories, when in the horizon crossing region, one can safely assume the following estimates to hold:  $\dot \pi \sim H \pi; \quad \nabla \pi \sim H/c_s\,\, \pi$ and so for $c_s\ll 1$ space-like derivatives dominate. We will provide a generalization of this argument (see Eq.~(\ref{estimates}) below). Let us stress already at this stage though that, should one decide to include all these subleading contributions in the action, the \textit{functional} expression of the solution will not change, one merely redefines a couple of time-independent coefficients. This is due to the fact that the types of operator ${\dot \pi}^2, (\partial_i \pi)^2,(\partial_i^2 \pi)^2 $ are already saturated at the level of Eq.~(\ref{l2}). 
\end{itemize}

\noindent We are now ready to tackle the equation of motion. After the usual change of variable,  $\pi(\vec k, t(\tau))=a(\tau) u(\vec k, \tau)$, the equation of motion can be written as:
\be
u'' -\frac{2}{\tau^2} u + \alpha_0 k^2 u + \beta_0 k^4 \tau^2 u=0 \label{eom},
\ee
where  $\alpha_0, \beta_0$ are time independent (again, at leading order) dimensionless coefficients. This equation has been written in the 
context of tilted Ghost Inflation~\cite{tilted} and to our knowledge, it has not been solved analitically before Ref.~\cite{b}, where the analytical solution has been briefly introduced and used for the computation of the three-point function. Here we discuss in much more details the 
properties of this solution. 
\noindent At this stage one can immediately recognize $\alpha_0$ as the more common $c_s^2$ and $\beta_0$ as the constant $\alpha^2 H^2 /M^2$ first introduced in \cite{ghost}. The complete expression for the coefficients is:
\be
\alpha_0 = \frac{-M_{Pl}^{2}\dot H - \bar M_{1}^{3}H}{-M_{Pl}^{2}\dot H +2M_{2}^{4} }; \qquad \beta_{0}=\frac{(\bar M_{2}^{2}+\bar M_{3}^{2})H^2}{2(-M_{Pl}^{2}\dot H +2M_{2}^{4})},
\ee
so that one reobtains the actual $c_s^2$ for $\bar M_{1}=0$. Note that one can simply look up the e.o.m. solution for DBI-like inflation if $\beta_0=0=\bar M_1$ and Ghost Inflation in the de Sitter limit provided $\alpha_0=0$. Let us pause here to comment on the possibility of a negative $\alpha_0$ (see also \cite{ssz05}). Such a scenario would result in a region in the $k$-space, whenever
\be | \alpha_0| k^2 \gg \beta_0 k^4 \tau^2 -\frac{2}{\tau^2},
\ee
for which the solution to the equation of motion will behave exponentially. Such a possibility raises a number of issues we address below. First of all, in order to keep control of the negative $\alpha_0$ region of the parameters space of the theory in the ultraviolet, one requires that the (positive) $\beta_0 k^4 \tau^2$ prevails over the $\alpha_0$ contribution before $k$ reaches the cutoff scale $\Lambda$. Considering that on the IR side, as we will show, the modes will eventually freeze outside the horizon, the case of a negative $\alpha_0$ should not in principle be disregarded. On the other hand, a lot of care should be exerted because an exponential phase of the modes for a sufficiently wide $k$ region could generate values for higher order correlators that directly contradict available observational data. 

\noindent We could now proceed to solve the complete equation of motion. Equipped with just the equation (\ref{eom}), we can already make some educated guesses on the behaviour of the wavefunction. First of all, the typical oscillatory behaviour deep inside the horizon is to be expected in this more general case as well: both $\alpha_0 k^2$ and $\beta_0 k^4 \tau^2$ cause wave-like behaviour (see Fig~1 below) of the wavefunction, while the $(-2/\tau^2)$ contribution is negligible. This is important in that it tells us the main contribution to correlation functions will be coming, as usual, from the horizon-crossing region. Note here that, as far as $\beta_0 \ne 0$, the 'Ghost Inflation' term will eventually lead the oscillation if one goes deep enough inside the horizon.\\
On the other hand, in the $\tau \rightarrow 0$ limit, $(-2/\tau^2)$ will be leading the dynamics and we expect to recover the usual, frozen modes. As is familiar from the DBI-like cases, it is convenient to introduce the notion of an effective horizon, placing it where the oscillatory behaviour stops being dominant. In formulas:
\be
 \alpha_0 k^2 + \beta_0 k^4 \tau_{*}^2 = \frac{2}{\tau_{*}^2} \quad \Rightarrow \quad \tau_{*} = -\frac{2}{k \sqrt{\alpha_0+ \sqrt{\alpha_0^2 + 8 \beta_0}}}\label{h}.
\ee
For $\beta_0=0, \quad \alpha_0\sim 1 $ one recovers $k^2 \tau_{*}^2 \sim 1$ at the horizon.\\
At this stage we can perform a consistency check and show how one can generalize the argument, initially borrowed from DBI-like inflationary models, that in comparing terms at the same order in perturbations and with the same overall number of derivatives, the ones with the most space derivatives are dominating in the $c_s \ll 1 $ limit. The generalization of this argument consists in restricting the parameters space to the $\alpha_0 \ll 1 \quad and \quad \beta_0 \ll 1$ region. Consider Eq.~(\ref{eom}) in Fourier space; in full generality one expects $\nabla \pi \sim k \pi$ and $\dot \pi \sim H \pi$ so what needs to be done is relate $k$ with $H$ at the horizon. Using equation (\ref{h}) and $\tau \sim -1/(a H)$ one obtains 
\be
k= \frac{\sqrt{2}H}{\sqrt{\alpha_0 + \sqrt{\alpha_0^2 + 8 \beta_0}}}. \label{estimates}
\ee
Since the main contributions to correlators comes from the horizon-crossing region, this shows that, for $(\alpha_0, \beta_0) \ll 1$ we can still identify leading terms in the Lagrangian according to the standard procedure.

\subsection{Wavefunction}
Let us verify all this quantitatively. The solution to Eq.~(\ref{eom}), being of second order, will come with two $k$-dependent integration constants. We have determined their values by requiring to re-obtain the known DBI and Ghost solutions in the corresponding limits. The general wavefunction reads:

\bea 
u_k(\tau)=\frac{i e^{\frac{1}{2} i \sqrt{\beta_0} k^2 \tau ^2}}{2^{1/4} \tau }\, \mathcal{G}\left[-\frac{1}{4}-\frac{i \alpha_0}{4 \sqrt{\beta_0}},-\frac{1}{2},-i \sqrt{\beta_0} k^2 \tau ^2 \right] C_1(k) 
\nonumber 
\eea
\bea
+ \frac{i e^{\frac{1}{2} i \sqrt{\beta_0} k^2 \tau ^2}}{2^{1/4} \tau } \mathcal{L}\left[\frac{1}{4}+\frac{i \alpha_0}{4 \sqrt{\beta_0}},-\frac{3}{2},-i \sqrt{\beta_0} k^2 \tau ^2\right] C_2(k)\,\, ,\nonumber\\ 
\eea
Where $\mathcal{G}$ stands for the confluent hypergeometric function and $\mathcal{L}$ is the generalized Laguerre polynomial.
We verified that, properly adjusting the integration constants according to
\be
\fl \qquad C_1(k)= \frac{  \left(\alpha_0+\sqrt{\beta_0}\right)^{-3/4}\Gamma\left[\frac{5}{4}-\frac{i \alpha_0}{4 \sqrt{\beta_0}}\right]k^{-3/2}}{ \sqrt{M_{Pl}^2 \epsilon\, H+2M_2^4}\,\, 2^{1/4}\, \Gamma\left[3/2-\frac{ \sqrt{\beta_0}}{4 \left(i \alpha_0+ \sqrt{\beta_0}\right)}\right]}\,\,;\qquad C_2(k)=0,
\ee
\noindent one obtains, in the appropriate limits \cite{b&d}, the wavefunctions of standard inflation and Ghost Inflation \cite{ghost}. We can now write our solution:

\be
\pi_k(\tau)=\frac{H\, e^{\frac{1}{2} i \sqrt{\beta_0} {k}^2 {\tau}^2} k^{-3/2} \Gamma (\frac{5}{4}-\frac{i \alpha_0}{4 \sqrt{\beta_0}}) \mathcal{G}(\alpha_0,\beta_0,k^2,{\tau^2})}{i\sqrt{M_{P}^2\epsilon H^2 +2M_2^4}\,  \sqrt{2}\, \gamma_0^{3/4}  \Gamma(\frac{5}{4}+\frac{\alpha_0}{4 \alpha_0-4 i \sqrt{\beta_0}})} \, , \label{sol}
\ee
where $\gamma =\alpha_0+\sqrt{\beta_0}$ and $\Gamma(x)$ is the Euler gamma function.\\
We see in particular that for $\alpha_0=0$, Eq.~(\ref{sol}) immediately reduces analitically to the Ghost Inflation wavefunction $\pi_k(\tau)=(H (-\tau)^{3/2}/\sqrt{2} M_2^2) \sqrt{\frac{\pi}{8}} \mathcal{H}_{3/4}^{1}(\frac{1}{2}\sqrt{\beta_0}k^2 \tau^2)$ with $\mathcal{H}_{3/4}^{1}$ being the Hankel function of the first kind. On the other hand, one can easily see numerically that the DBI solution is recovered in the $\beta_0 \rightarrow 0$ limit.\\ To give some intuition on the behaviour of the general, interpolating wavefunction, we plot it in several $(\alpha_0, \beta_0)$ configurations. For overall consistency in the comparisons, in all the following pictures we have chosen points in the $(\alpha_0, \beta_0)$-plane so that the horizon crossing always lies at the same point, numerically $\tau_{*}=-\sqrt{2}$, and we have plotted the wavefunction from well inside the horizon ($\tau=-10\,\, \tau_{*}$) up to $\tau=0$.\\
\newpage
\begin{figure}[hp]
	\centering
		\includegraphics[width=0.43\textwidth]{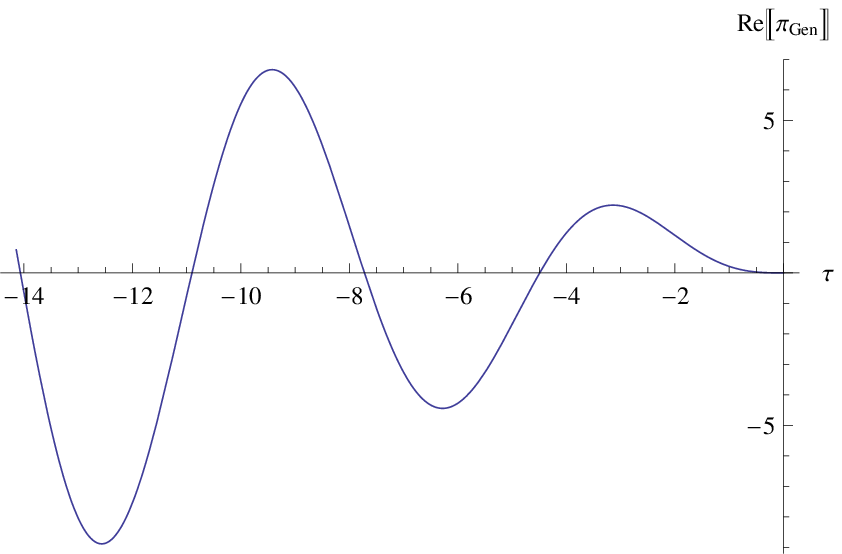}
		\hspace{8mm}
			\includegraphics[width=0.43\textwidth]{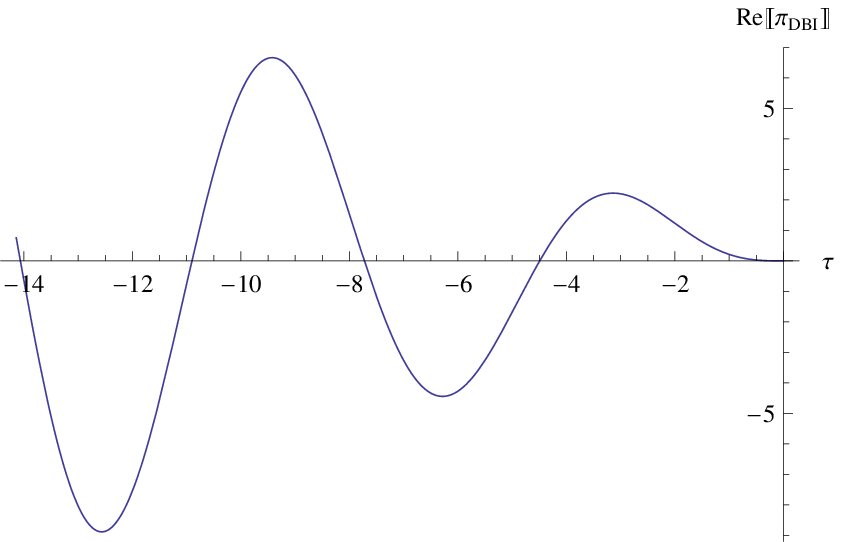}
	\caption{On the left, the general wavefunction in the DBI-like, $\beta_{0}\rightarrow 0$, limit; on the right the DBI-like solution itself. We plot the real part and find perfect agreement, same holds for the imaginary part. To produce the plot the parameters have been set to: $\alpha_0=0.1, \beta_0 \rightarrow 0, k=1, H=1$ and the Planck mass-dependent normalization has been neglected. The corresponding plot can be omitted for the Ghost limit since in that case we recover the Ghost solution analytically.}
	\label{fig:gen_dbi_limit}
\end{figure}

\begin{figure}[hp]
	\centering
		\includegraphics[width=0.43\textwidth]{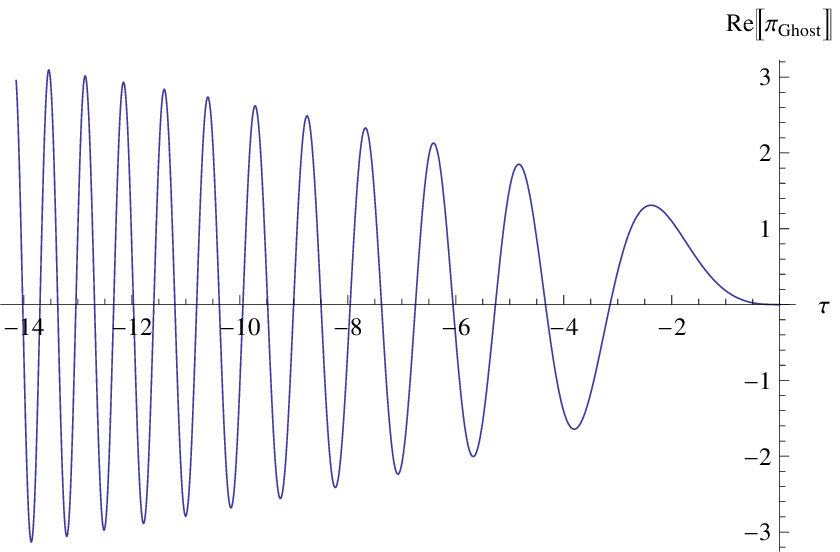}
		\hspace{8mm}
			\includegraphics[width=0.43\textwidth]{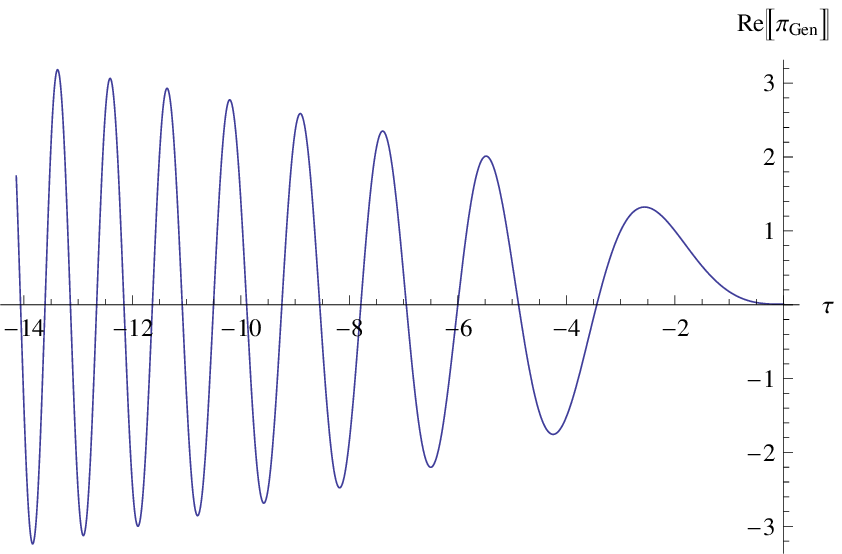}
	\caption{On the left, the Ghost Inflation wavefunction ($\alpha_0=0$). On the right the general interpolating solution calculated for $\alpha_0=1/2\,,\, \beta_0=1/4$.}
	\label{ghostandgen}
\end{figure}

\noindent From these plots we learn several things. First, as argued before, there is a common oscillatory behaviour once inside the horizon. The frequency of these oscillations is more pronounced for the Ghost solution when deeper inside the horizon. In the general solution the frequency varies according to the ``relative weight'' of the Ghost component, $\beta_0$, and the DBI-like one, $\alpha_0$.
\subsection{Power spectrum}
We now turn to the expression of the power spectrum

\bea
\fl P_{\pi}= \frac{k^3}{2 \pi ^2}|\pi(k,\tau \rightarrow 0)|^2=\frac{H^2}{16 \pi  (M_{P}^2 \epsilon H^2 +2M_2^4) (\alpha_0 + \sqrt{\beta_0})^{3/2}\,
|\Gamma(\frac{5}{4}+\frac{\alpha_0}{4 \alpha_0-4 i \sqrt{\beta_0}})|^2} .\nonumber\\
\eea
Clearly there is no time dependence in the above result, the modes freeze outside the horizon. Note here that, to reproduce standard results, we should re-introduce the speed of sound $c_s$ which, in the language we are using, is related to $M_2$ via:
\[
\frac{1}{c_s^2}=1-\frac{2 M_2^4}{M_P^2 \dot H}.
\] 
Upon switching to the gauge invariant quantity related to $\pi$ by $\zeta=-H \pi $ \cite{eft08} (see also the appendix of \cite{3pt} for the relation at second order), and 
reintroducing the proper units with Planck mass we get:
\be
P_\zeta=\frac{ (\alpha_0 + \sqrt{\beta_0})^{-3/2}  H^4}{16 \pi   (M_{Pl}^2 \epsilon H^{2}+2M_2^4) \,
|\Gamma(\frac{5}{4}+\frac{\alpha_0}{4 \alpha_0-4 i \sqrt{\beta_0}})|^2}. \label{p1}
\ee
Again, one could easily check that the above result analytically covers the power spectrum of DBI-like and Ghost Inflation.\\
As we mentioned in the last comments to equation (\ref{l2}), even when including subleading terms in the quadratic action, the functional dependence of our wavefunction does not change, only the definition of $\alpha_0, \beta_0$ does. Since we are now going to set bounds on operators coefficients , we want to be as precise as possible and will therefore extend the definition of the two parameters to cover the subleading terms as well. We now have: 
\bea
\fl \qquad \alpha_0 = \frac{-M_{Pl}^{2}\dot H - \bar M_{1}^{3}H/2 }{-M_{Pl}^{2}\dot H +2M_{2}^{4} - \bar 3 \bar M_{1}^{3}H }\,\, ; \qquad \beta_{0}=\frac{\bar M_{0}^{2}H^2/2}{-M_{Pl}^{2}\dot H +2M_{2}^{4} - 3\bar M_{1}^{3}H}. \label{param}
\eea

\noindent In obtaining Eq.~(\ref{param}), we took into account the fact that the $\bar M_{1},\bar M_{2},\bar M_{3}$-driven terms multiply operators of the type  ${\dot \pi}^2, (\partial_i \pi)^2$ as well. We also choose to replace $\bar M_{2}$ and $\bar M_{3}$ with a linear combination of the two masses: we set $\bar M_3^2 =-3\bar M_2^2$ and $\bar M_0^2 =\bar M_2^2+\bar M_3^2$, see also \cite{b}. This procedure allows one to put to zero all the subleading operators tuned by $\bar M_{2},\bar M_{3}$ and makes the correspondence between inflationary models and the switching of the $M,\bar M$ parameters absolutely sharp \footnote{The reader might worry that one degree of freedom is lost. However, the two coefficients multiply basically the same interaction terms in the action up to
fourth order.}.\\ An immediate simplification is that now, upon requiring $\bar M_0=0 \Leftrightarrow \beta_0 =0$, one goes into DBI inflation, \textit{exactly}. Similarly, now de-Sitter limit and $\bar M_1=0$ give Ghost Inflation with $\alpha_0=0$. The power spectrum looks very similar to the one in Eq.~(\ref{p1})

\be
P_\zeta=\frac{ (\alpha_0 + \sqrt{\beta_0})^{-3/2}  H^4}{16 \pi   (M_{Pl}^2 \epsilon H^{2}+2M_2^4-3 \bar M_1^3 H_{}) \,
|\Gamma(\frac{5}{4}+\frac{\alpha_0}{4 \alpha_0-4 i \sqrt{\beta_0}})|^2}, \label{p2}
\ee
only the definitions of the $\alpha_0,\beta_0$ parameters and the normalization constant have slightly changed. Expressing all the parameters in the spectrum in terms of the $M, \bar M$ coefficients, we count the degrees of freedom as being five, associated to $H, \epsilon, M_2, \bar M_0, \bar M_1$. The first three are the same that appear also in the standard case as $H, \epsilon, c_s$; using $c_s$ instead of $M_2$ is just a matter of dictionary. In the Ghost Inflation case, the quantity $\bar M_0$ replaces the speed of sound and we are again back to three parameters. In the most general case one has to keep both $\bar M_0$ and $\bar M_1$ as well. Bounds can be put on the values of these five parameters by employing the expected value for the power spectrum, $P_{k}^{\zeta}\sim 10^{-10}$ and its tilt. Let us also mention here that further mild inequalities must be satisfied by the $\alpha_0, \beta_0$ parameters in order to keep the generalized speed of sound small, see \cite{b}. In \textit{Sec. 2.3} below we present a calculation for the tilt and running of the power spectrum. These quantities are essentially obtained from considering the time dependence of the $M$ coefficients in Eq.~(\ref{p2}). On the other hand, as has been specified above, part of the procedure that led to Eq.~(\ref{p2}) has been to disregard the time dependence of said coefficients \footnote{This is because, at the level of the action, considering the time dependence of these coefficients would automatically translate into going at next-to-leading order in slow-roll; for a leading-order calculation it is therefore sufficient to consider their values at the horizon.}. Restoring it at a later step, as we do below, is standard accepted procedure because generally only at this stage the effect of time-dependence becomes important. A calculation that does without this assumption is presented in \textit{Sec. 3}.
\subsection{Tilt and running}
Below we employ some simplification in order to present our result for the tilt of the power spectrum in a way that resembles as closely as possible the typical expression for  $n_s - 1$. Indeed, the spectrum dependence on the Euler $\Gamma$ function in Eq.~(\ref{p2}) is not to be found in e.g. DBI, Ghost Inflation etc. For simplicity, we choose not to write here the explicit dependence of the Euler function on the $M$ coefficients and leave it implicit; we report the full explicit dependence in the \textit{Appendix}.
We apply the following formula,
\be
\fl
n_s-1=  \frac{d}{d \ln{k}}\ln{P_k}=\left(\frac{d \ln{k}}{dt}\Big{|}_{t=t^*}  \right)^{-1}\frac{1}{P_k}\frac{d P_k}{dt}\Big{|}_{t=t^*} \simeq\frac{1}{H P_k}\frac{d P_k}{dt}\Big{|}_{t=t^*} ,
\ee
on the power spectrum, where $t^{*}$ is the time at horizon crossing. The time dependence of the $M, \bar M$ coefficients is taken into account and the time dependence of the Euler function is dealt with as one would do with a generic function  $\Gamma(t)$. One obtains:
\bea
n_s-1= -\frac{\dot \Gamma}{H \Gamma}-\epsilon\times \frac{7 H^2 M_P^2 \epsilon+8 \left(2 M_2{}^4+3 M_1{}^4\right) }{2  \left(H^2 M_P^2 \epsilon+2 M_2{}^4+3 M_1{}^4\right)}+\nonumber\\ 
\fl -\epsilon \times  \frac{8 H^4 M_P^4 \epsilon^2-16 H^2 M_P^2 \epsilon \left(2 M_2{}^4+3 M_1{}^4\right)+H^2 M_P^2 \epsilon  \left(-14 H^2 M_P^2 \epsilon +8 M_2{}^4+15 M_1{}^4\right) }{2  \left(H^2 M_P^2 \epsilon+2 M_2{}^4+3 M_1{}^4\right) \left(2 H^2 M_P^2 \epsilon+M_1{}^4+M_0{}^2 \sqrt{2 H^2 M_P^2 \epsilon +4 M_2{}^4+6 M_1{}^4}\right)}\nonumber\\
\fl- \eta \times \frac{ \left(6 H^2 M_P^2 \epsilon +24 M_2{}^4+33 M_1{}^4\right)H^2 M_P^2 \epsilon }{4 \left(H^2 M_P^2 \epsilon +2 M_2{}^4+3 M_1{}^4\right) \left(2 H^2 M_P^2 \epsilon +M_1{}^4+M_0{}^2 \sqrt{2 H^2 M_P^2 \epsilon +4 M_2{}^4+6 M_1{}^4}\right)}\nonumber\\
\fl - \eta \times \frac{ H^2 M_P^2  \epsilon }{4 \left(H^2 M_P^2 \epsilon +2 M_2{}^4+3 M_1{}^4\right)}-\frac{\dot M_2}{H M_2}\times \frac{ 2 M_2{}^4}{ \left(H^2 M_P^2 \epsilon +2 M_2{}^4+3 M_1{}^4\right)}+\nonumber\\
\fl+ \frac{\dot M_2}{H M_2}\times \frac{ \left(6 H^2 M_P^2 \epsilon +3 M_1{}^4\right)2 M_2{}^4 }{ \left(H^2 M_P^2 \epsilon +2 M_2{}^4+3 M_1{}^4\right) \left(2 H^2 M_P^2 \epsilon +M_1{}^4+M_0{}^2 \sqrt{2 H^2 M_P^2 \epsilon +4 M_2{}^4+6 M_1{}^4}\right)}\nonumber \\
\fl -\frac{\dot M_1}{H M_1}\times \frac{\left(+ 3M_1{}^4-4 H^2 M_P^2 \epsilon +4 M_2{}^4\right)3 M_1{}^4 }{ \left(H^2 M_P^2 \epsilon +2 M_2{}^4+3 M_1{}^4\right) \left(2 H^2 M_P^2 \epsilon +M_1{}^4+M_0{}^2 \sqrt{2 H^2 M_P^2 \epsilon +4 M_2{}^4+6 M_1{}^4}\right)}\nonumber \\ 
\fl  -\frac{\dot M_0}{H M_0}\times \frac{6 M_0^2 }{ \left(2 M_0{}^2+\sqrt{2} \left(2 H^2 M_P^2 \epsilon +M_1{}^4\right) \sqrt{\frac{1}{H^2 M_P^2 \epsilon +2 M_2{}^4+3 M_1{}^4}}\right)} \nonumber \\
\fl -\frac{\dot M_1}{H M_1}\times \frac{3 M_1{}^4 }{ \left(H^2 M_P^2 \epsilon +2 M_2{}^4+3 M_1{}^4\right) } \, ,  \label{ns}
\eea
where all the quantities are to be intended as calculated at horizon crossing.\\
For convenience, we have factored out the usual parameters: $\epsilon$, $\eta$, $s$ (the latter is written in $M_2$ language; we give the dictionary in Eq.~(\ref{e2}) below) and their generalization:
\bea
\epsilon_{0}=\frac{\dot M_0}{H M_0}\,\,; \qquad \epsilon_{1}=\frac{\dot M_1}{H M_1}\,\,; \qquad \epsilon_{\Gamma}=\frac{\dot \Gamma}{H \Gamma}\,\,. \label{genep}
\eea
\bea
 \epsilon_2\equiv \frac{\dot M_2}{H M_2}=\frac{\eta}{2}-\epsilon +\frac{s}{2(c_s^2-1)} \label{e2}\,. 
\eea
\noindent The variable $c_s$ is the one defined in \textit{Sec. 2.1} and $\eta=\dot \epsilon /(H \epsilon)$. For simplicity, we have defined the variables $M_0^4 \equiv \bar M_0^2 H^2$ and $M_1^4 \equiv \bar M_1^3 H$.\\
\noindent Let us briefly comment on the above results. First note that, as expected, the factor that each of these generalized slow-roll parameter multiplies is of order unity or smaller. A quick consistency check consists in specializing the formula above to known inflationary models, for example requiring $\bar M_1=0=\bar M_0$ gives back the usual result \cite{3pt} for DBI-like models:
\bea
n_s-1= -2\epsilon -\eta -s
\eea
Consider now the more general case with $(M_1=0\,,\dot H \not= 0\,,M_0\not= 0\,) ,$ which comprises  DBI-like theories and Ghost Inflation models as limiting cases. In such a scenario there are some mild bounds to be required on $M_0^4, M_{P}^2 \epsilon H^2$. First of all, since we are interested in the $\alpha_0 \ll 1\, , \beta_0 \ll 1$ region of the parameters space this requires:
\bea
M_{P}^2 \epsilon H^2 \ll M_2^4\,\,;  M_0^4 \ll M_2^4 \,.
\eea
On the other hand, the $M_0$-driven slow-roll parameters is $\epsilon_0= \dot M_0 /(H M_0)$ and therefore the above inequalities do not put upper bounds on this slow-roll parameter. Much like we will see in the next section for the running of the bispectrum amplitude, one must instead be careful to account for the fact that too large a value for $\dot M_0$ could give a contribution to the bispectrum amplitude that must be excluded. Indeed, once expanded, the $M_0$-proportional contribution to the quadratic action for the scalar reads:
\bea
\fl \int{ d^4 x \sqrt{-g}\Big[..+ M_0^4(t+\pi) \frac{(\partial_i^2 \pi)^2}{a^4}   \Big]}\sim  \int{ d^4 x \sqrt{-g}\Big[..+ M_0^4(t) \frac{(\partial_i^2 \pi)^2}{a^4}+ 4 M_0^3(t) \dot M_0 \pi\frac{(\partial_i^2 \pi)^2}{a^4}   \Big]},\nonumber\\
\eea
\noindent with the second term on the RHS action clearly contributing to the three-point function of the scalar.  This $\dot M_0$-generated contribution must be weighted against the third order interactions generated by $M_0$ itself, but also by new $M$ coefficients that first appear at third order in the action (see \cite{b} and the analysis in \textit{Sec. 4}). A similar analysis applies for $\epsilon_1$ if we let $M_1 \not= 0$.\\
We can conclude that the mild bounds on these new, generalized slow-roll parameters come from the obvious fact that we are doing a slow-roll expansion, from the value of the power spectrum itself and from the requirement that they are not so large as to produce too large a value for $f_{NL}$. \\
In order to obtain the running of the power spectrum we proceed as below:
\be
\alpha_{s}= \frac{d n_s}{d\log k} \simeq \frac{1}{H}\frac{d n_s}{d t}.
\ee
We give below a compact results: 
\bea
\fl \alpha_s= -\frac{\ddot \Gamma}{H^2 \Gamma} + \epsilon_{\Gamma}^{2}- \frac{\dot \epsilon \Theta_{}}{H }- \frac{ \epsilon \dot \Theta_{}}{H }- \frac{\dot \epsilon_{2} \Theta_{2}}{H }- \frac{ \epsilon_{2} \dot \Theta_{2}}{H }
- \frac{\dot \eta \Theta_{\eta}}{H }- \frac{ \eta \dot \Theta_{\eta}}{H }
- \frac{\dot \epsilon_{1} \Theta_{1}}{H }- \frac{ \epsilon_{1} \dot \Theta_{1}}{H }
- \frac{\dot \epsilon_{0} \Theta_{0}}{H }- \frac{ \epsilon_{0} \dot \Theta_{0}}{H },\nonumber\\ \label{run}
\eea
and point the reader to the \textit{Appendix} for a more explicit expression of the coefficients functions $\Theta,\Theta_{2},\Theta_{\eta},\Theta_{1},\Theta_{0}$.
\section{Next-to-leading order}
The discussion presented so far is based on a generalized slow-roll approximation at leading order. In particular, the $M , \bar M$ coefficients driving the various operators in the Lagrangian are assumed to be time independent. This assumption propagates into the equation of motion for the scalar $\pi$, the classical solution itself and the power spectrum. To make up for this approximation when calculating the the tilt of the spectrum, one restores the time dependence of the coefficients at the level of the power spectrum. A more systematic approach consists in accounting for the time dependence of the coefficients already at the Lagrangian level, this is done by taking the generalized slow-roll approximation to next order. Schematically one has:
\bea
\fl S^{\pi}_{2}\propto \int d^4 x \sqrt{-g} \Big[ -M_P^2 (3H^2(t+\pi)+\dot H(t+\pi))+M(t+\pi)\times (\rm{ quadratic})   \Big] \, . \label{2nd}
\eea
\noindent The first term does not appear in the Lagrangian in Eq.~(\ref{l2}) because, at leading order in slow roll, it is not quadratic in fluctuations, it does in fact, contribute to the background. We see that accounting for the ``$\pi$'' in the time dependence of $M$ in the last term in Eq.~(\ref{2nd}) would result in a third order operator. This must be considered when studying interactions but it is not what we want to analyze here, the wavefunction comes from the quadratic Lagrangian. On the other hand, the ``$\pi$'' in the first term of the action has to be accounted for; doing so results in just one additional contribution to the action and it turns out  to be proportional to $\epsilon ^2$. This is all consistent with the fact that, at leading order, the action is instead proportional to $\epsilon$. We now have:
\bea
\fl \mathcal{L}_2=a^3&\Big[&M_{P}^{2}\dot H(t) (\partial_{\mu} \pi)^2 
+ 2 M_2^4(t){\dot\pi}^2 - \bar M_1^3(t) H(t) (3 {\dot \pi}^2-\frac{(\partial_i \pi)^2 }{2 a^2})  \nonumber\\
\fl &+& \frac{\bar M_0^2(t)}{2}\frac{1}{a^4} (\partial_i^2\pi)^2 -3 M_P^2 {\dot H(t)}^2 \pi^2  \Big] \, .
\label{l22}
\eea
We proceed to write down the equation of motion as in the leading order case, obtaining:
\bea
\fl \sigma_k^{''}+\alpha_0(1+s_{\alpha})k^2 \sigma_k+\beta_0(1+s_{\beta}) \frac{k^4}{a^2 H^2}\sigma_k = \left(\frac{f^{''}}{f}+3\epsilon a^2 H^2\frac{ M_P^2 {\dot H(t)}}{2M_2^4-M_P^2 \dot H -\bar M_1^3 H}   \right)\sigma_k\, ,\nonumber\\ \label{eoma}
\eea
where the following definitions have been employed:
\bea
\fl s_{\alpha}=\frac{\dot \alpha_0}{ H \alpha_0}\ll 1\,; \qquad s_{\beta}=\frac{\dot \beta_0}{ H \beta_0}\ll 1\,;\quad f^2=a^2( -M_P^2 \dot H +2 M_2^4 -3\bar M_1^3 H)\,; \quad \pi=\frac{\sigma}{f}. \nonumber \\
\eea
In order to solve the equation of motion one needs to calculate $f^{''}/f$ explicitly; the result is given in the Appendix.
A compact expression for the equation of motion is given by:
\bea
 \sigma_k^{''}+\tilde \alpha_0 k^2 \sigma_k+\tilde \beta_0 {k^4 \tau^2}\sigma_k =\frac{2}{\tau^2}(1+x_0)\sigma_k\, \label{eomb} \, ,
\eea
where $x_0$ is a linear combination of slow roll parameters and $\tilde \alpha_0,\tilde \beta_0$ represent a slight redefinition of the initial parameters. For explicit expression we refer once again the reader to the Appendix. Equipped with Eq.~(\ref{eomb}), one uses the Bunch-Davies vacuum condition to write down the solution:
\bea
\fl \qquad \sigma(x,k)=C_1 (k) \frac{ \mathcal{G}\left[\frac{-i \tilde\alpha_0 \sqrt{\tilde \beta_0}+2 \tilde \beta_0 + \tilde \beta_0 \sqrt{9+8 x_0}}{4 \tilde\beta_0},\frac{1}{2} \left(2+\sqrt{9+8 x_0}\right),-i \sqrt{\tilde \beta_0} k^2 x^2\right]}{2^{-\frac{1}{4} \left(2+\sqrt{9+8 x_0}\right)} e^{-\frac{1}{2} i \sqrt{\tilde \beta_0} k^2 x^2} \left(x^2\right)^{\frac{1}{4} \left(2+\sqrt{9+8 x_0}\right)}\sqrt{x}} \, ,
\eea
where $\mathcal{G}$ is the usual hypergeometric function, $x = -\tau$, and the wavefunction must be expanded to first order in $x_0 = 0$. In the first section, we gave the exact expression for the solution above at leading order and went on to calculate the resulting power spectrum, its tilt and running. The calculation at next-to-leading order has already been performed for DBI-like theories of inflation, using the same formalism employed here, in \cite{3pt}. Here instead, we choose to calculate the next-to-leading order Ghost Inflation solution obtaining also the tilt of the power spectrum and the running. The procedure is a standard one, so we briefly sketch it. The two $k$-dependent constant of the Ghost equation of motion
\bea
 \sigma_k^{''}+\tilde \beta\, {k^4 \tau^2}\sigma_k =\frac{2}{\tau^2}(1+x_0^{G})\sigma_k\,, \label{eomghost}
\eea
are reduced to one by imposing the correct leading order limit on the wavefunction. The remaining constant is obtained by requiring the proper normalization, that is by imposing the following commutation relations to hold:
\bea
[\pi(\vec x),P(\vec y)]=i\, \delta^3(\vec x -\vec y)\,; \qquad [a_{\vec k},a^{\dagger}_{\vec p}]= (2\pi)^3 \delta^3(\vec k -\vec p) \, ,
\eea
where $P$ is the momentum conjugate of the scalar $\pi$ and the creation and annihilation operators $a,a^{\dagger}$ are the usual operators in the free field expansion for the quantized field $\pi$. Proceeding as prescribed above, one obtains:
\bea
\fl \pi(k,\tau)= \frac{H \sqrt{2 \pi } (1+s_{\beta})^{1/8} (-\tau )^{3/2} \, H^{(1)}\left[\frac{1}{4} \sqrt{9+8 x_0},\frac{1}{2} k^2 \sqrt{\beta_0 (1+s_{\beta})} \tau^2\right]}{3 \left(1+\frac{s_{\beta}}{8}-x_0 (-\frac{5}{9}+\gamma/3-\frac{\pi }{6}+\log{2})\right)\left( \Gamma \left[1+\frac{1}{4} \sqrt{9+8 x_0}\right]\right)^{-1} \Gamma \left[\frac{3}{4}\right]}.\label{ghostsol}
\eea
The wavefunction above is the correct Ghost Inflation wavefunction up to next-to-leading order in generalized slow-roll parameters; it gives back the leading order solution and respects the proper Bunch-Davies vacuum requirement. From Eq.~(\ref{ghostsol}) one can readily calculate the tilt of the spectrum: one simply considers the leading behaviour of the wavefunction as $k$ goes to zero. In fact, upon expanding for small $k$ one finds that our solution goes like:
\bea
 H^{(1)}\left[\frac{3}{4}+\frac{x_0}{3},\frac{1}{2} \sqrt{\beta_0}\left(1+\frac{s_{\beta}}{2}\right) k^2 \tau^2\right]\sim k^{-\frac{3}{2}-\frac{2 x_0}{3}} \, ,
\eea
from which we obtain that 
\bea n_s-1 = -4/3 x_0.
\eea 
\noindent When specialized to Ghost Inflation, the value of the $x_0$ parameter (which always constists of a linear combination of the generalized slow roll parameters) is given by:
\bea
x^{G}_0=3 \frac{\dot M_2}{H M_2},
\eea
and the running amounts to simply 
\bea \frac{d n_s}{d\ln{k}}= -4\, \dot{x_0}/3H .
\eea

\section{Running of $f_{NL}$}
In the quest for properties that help in removing degeneracies among the many inflationary models one generally considers another observable beyond the power spectrum and its running,  i.e. the analysis of non-Gaussianities. Starting with the bispectrum, one can study its amplitude, shape and running. In the same spirit of the analysis we performed for the power spectrum, we now want to estimate the value for the running of the bispectrum amplitude, $f_{NL}$. In performing the calculation for $P_{\zeta}$, $n_s-1$ and then $\alpha_s$, we used the fact that the coefficients driving quadratic operators in the Lagrangian are nearly constant, up to slow-roll corrections. At first approximation, one writes down the power spectrum as a function of these parameters calculated at the horizon. Only when calculating the tilt of the spectrum and its running one does consider the time dependence on the $M$'s, thus obtaining Eq.~(\ref{ns}),(\ref{run}). Similarly here, we will employ the results on the bispectrum amplitude contributions generated by independent interaction terms as given in \cite{b} (see \cite{in-in1}-\cite{w-qccc} for the \textit{in-in} formalism). In particular we are going to focus on the running of $f_{NL}$ generated by a third-order interaction term whose bispectrum shape-function peaks is an uncommon flat configuration (see also \cite{ssz05}). In fact, in \cite{b} there are several independent terms that generate at least two qualitatively different flat shape-functions; we choose here to concentrate on the analysis of just one term as the same considerations and conclusions can be straighforwardly adapted to all of them.\\
The analysis for the running of $f_{ NL}$ has been done for several inflationary mechanisms such as DBI inflation, and others. We are going to extend this type of study to a third order interaction term driven by a nearly constant coefficient, $\bar M_6(t)$, and which is generated by an extrinsic curvature contribution: \footnote{The contribution of $\bar M_6$ to the action at third order is proportional to $-\bar M_6^2(g^{00}+1)\delta K_{\mu}^{\nu} \delta K_{\nu}^{\mu}$ \cite{b}, where $K_{\mu\nu}$ is the extrinsic curvature tensor.}
\be
\int d^3x dt \sqrt{-g}\Big[...+\frac{\bar M_6(t+\pi)^2}{3} \frac{\dot \pi}{a^4} \sum_{i,j}(\partial_{ij}\pi)^2+...\Big].\label{m6}
\ee
In \cite{b} we calculated the corresponding contribution to $f_{NL}$ in six different points of the $(\alpha_0,\beta_0)$ plane, we report them in \textit{Table 1}.
\begin{center}
\begin{tabular}{| l || l | l | l | l | l| l | }
\hline			
       Benchmarks   & 1 & 2 &3 &4 &5 &6 \\ \hline 
  $\alpha_0$ & $10^{-2}$  & 0 & $0.5\cdot 10^{-2}$ & $2 \cdot 10^{-7}$ &$10^{-4}$  &$10^{-6}$ \\ \hline
  $\beta_0$  &  0 & $0.5\cdot 10^{-4} $ & $0.25\cdot 10^{-4} $ & $5\cdot 10^{-5}$ & 0&0  \\
\hline  
\end{tabular}
\end{center}
\noindent{\scriptsize \textit{Table 1}: configuration \textit{1} describes pure DBI-like theories, pure ghost corresponds to configuration \textit{2}. In \textit{3,4} a more general model is considered while in the last two configurations one aims at considering the cases characterized by very small generalized speed of sound, $\sqrt{\alpha_0}$.
}\\

\vspace{2mm}
\noindent Corresponding to each one of the configurations of \textit{Table 1}, we report in \textit{Table 2} below the value of $f_{NL}$.
\begin{center}
\begin{tabular}{| l || l | l | l | l | l| l | }
\hline			
benchmarks   & 1 & 2 &3 &4 &5 &6 \\ \hline 
$f^{\bar M_6}_{ NL}$   &     $\,\,\,  10^{4}\, \gamma_6 \,\,\, $     &    $\,\,\, 4\cdot  10^{3}\,\gamma_6\,\,\, $       &     $\,\,\,5\cdot 10^{3}\,\gamma_6\,\,\,$     &      $\,\,\, 4\cdot 10^{3}\,\gamma_6\,\,\,$      &      $\,\,\, 10^8\,\gamma_6\,\,\,$    &    $\,\,\, 10^{12}\,\gamma_6\,\,\,$  \\ 
\hline  
\end{tabular}
\end{center}
\noindent{\scriptsize \textbf{ \textit{Table 2}: the value of the bispectrum amplitude $f^{\bar M_6}_{ NL}$ in the six different configurations.}
}
\vspace{5mm}

\noindent In \textit{Table 2} we introduced the dimensionless quantity $\gamma_6$, which is given by
\be
\gamma_6 = (\bar M_6^2 H^2 )/(M_{ Pl}^2 \epsilon H^2 + 2 M_2^4 -3\bar M_1^3 H).
\ee 
Computing the running of $f_{NL}$ amounts then to calculating
\bea
n_{NG}\equiv \frac{d\, \ln{|f_{NL}(k)|}}{d \ln{k}}\simeq \frac{1}{H f_{ NL}}\frac{d\,f_{ NL}}{dt}.
\eea
For the specific case at hand, we have:
\bea
\fl \frac{1}{H f_{NL}^{\bar M_6}}\frac{d}{dt} f_{NL}^{\bar M_6}=\frac{1}{H f_{NL}^{\bar M_6}}\frac{d}{dt}\Big[  N(t) \gamma_6(t)\Big]\Big{|}_{t=t^{*}}=
   -\epsilon \frac{4 M_2{}^4 +2 M_1{}^4}{\left(H^2 M_P^2 \epsilon +2 M_2{}^4+3 M_1{}^4\right)}+2 \epsilon_6+\epsilon_{N}+\nonumber\\
\fl -\eta \frac{H^2 M_P^2\epsilon  }{H^2 M_P^2 \epsilon +2 M_2{}^4+3 M_1{}^4}-\epsilon_2\frac{8 M_2{}^4 }{ \left(H^2 M_P^2 \epsilon +2 M_2{}^4+3 M_1{}^4\right)}-\epsilon_1\frac{4 M_1{}^4 }{ \left(H^2 M_P^2 \epsilon +2 M_2{}^4+3 M_1{}^4\right)}.   \nonumber\\ \label{f6}
\eea
In the above expression $N$ stands for a numerical factor which is dependent on $\alpha_0$ and $\beta_0$, specifically it goes like $(\alpha_0+\beta_0)^{-n}$ with $n$ small positive integer. The generalized slow-roll parameters are defined as usual, with the only new ones being $\epsilon_6$,$\epsilon_{N}$:
\bea
\epsilon_6 = \frac{\dot{\bar M_6}}{H \bar M_6}\,\,; \qquad \epsilon_{N}={\cal O}(1)\times (\epsilon,\eta,\epsilon_{2},\epsilon_{1},\epsilon_{0})\label{e6}.
\eea
The last expression in Eq.~(\ref{e6}) means that $\epsilon_{N}$ can be expressed as a linear combination of the generalized slow roll parameters introduced before multiplied at most by an order unity constant. Overall, we can then conclude that the running of $f_{NL}^{\bar M_6}$ is of the order of the generalized slow roll parameters (or a linear combination thereof). Now, all these parameters, except for $\epsilon_6$, can be expressed as functions of $M,\bar M$ coefficients that first appear in the quadratic Lagrangian of the theory. As such, four of these parameters could in principle be expressed as a function of the observables one usually uses, namely the scalar and tensor spectral indices and their running. Not so for $\epsilon_6$, on which, in principle, we enjoy more freedom as it drives terms in the action that are at least cubic. It could indeed be that the running of $f_{NL}^{\bar M_6}$ is dominated by this contribution and could therefore be larger than one finds in some general single-field slow-roll models where the extrinsic curvature-generated interaction terms are not accounted for. The same considerations apply to other interaction terms such as the $\bar M_9$-driven one in \cite{b}. On the other hand, special care must be exerted so as to make sure that requiring $\epsilon_6$ to be the leading generalized slow-roll parameter in Eq.~(\ref{f6})
does not spoil the possibility to have the contribution in Eq.~(\ref{m6}) dominate the overall bispectrum amplitude, which is what made this type of contribution interesting in the first place. Indeed, there are two ways of making $\epsilon_6$ the leading parameter, a large $\dot{\bar M_6}$ and a small $\bar M_6$. Pushing the latter option too far the interesting and possibly leading bispectrum and trispectrum amplitudes~\cite{b} would become subdominant and this would make the corresponding flat shape-functions a mere curiosity. But also the former option has to be discussed and this is clear from the expansion of Eq.~(\ref{m6}) resulting from considering the $\bar M_6$ time dependence:

\bea
\fl \int d^3x dt \sqrt{-g}\Big[...+\frac{\bar M_6(t)^2}{3} \frac{\dot \pi}{a^4} \sum_{i,j}(\partial_{ij}\pi)^2+\frac{ 2\bar M_6(t)\dot{\bar M_6}(t)}{3} \frac{\dot \pi}{a^4} \sum_{i,j}(\partial_{ij}\pi)^2 \times \pi...\Big].
\eea
Indeed the quartic interaction term that appears is proportional to $\dot{\bar M_6}$ and if the latter is too big it could give rise to too large a contribution to the power spectrum at one loop and would have to be ruled out. Below we give a number of inequalities that the quantities $\dot{\bar M_6}$, $\bar M_6$ need to satisfy in order not to spoil the appealing bispectrum and trispectrum features outlined above. It turns out they are not too restrictive and that the running of $f_{NL}^{\bar M_6}$ can be safely ruled by $\epsilon_6$. We first write down the inequalities that stem from requiring that the $\dot{\bar M_6}$-proportional quartic interaction term is not the leading interaction in the fourth order Lagrangian (this would have consequences on the loop corrections to the power spectrum as well) as compared to the usual single-field interactions (for the complete action at fourth-order see Ref.~\cite{t})
\bea
\fl \frac{M_2^4 }{\bar M_6^2 H^2}> \epsilon_6 \gg \epsilon\,; \qquad \frac{(\alpha_0+\sqrt{\beta_0}) M_3^4 }{\bar M_6^2 H^2}> \epsilon_6 \gg \epsilon\,;\qquad \frac{(\alpha_0+\sqrt{\beta_0})^2 M_4^4 }{\bar M_6^2 H^2} > \epsilon_6 \gg \epsilon\,.\label{ineq1}
\eea
Notice that only one of these inequalities need be satisfied. The $\epsilon_6 \gg \epsilon$ part of the inequalities above ensures that indeed $\epsilon_6$ is the dominating generalized slow-roll parameter.  One has to keep in mind here that large non-Gaussinities are generated by requiring the generalized speed of sound, $\sim \alpha_0+\sqrt{\beta_0}$, to be much smaller than unity and so the first of these inequalities seems somewhat less stringent than the others though there is no requirement on the $M$'s to be all of the same order.\\
As anticipated a small $\bar M_6$ can in principle lead to a subleading contribution to the bispectrum signal thus rendering the corresponding flat shape-function less interesting. Borrowing the third order action of \cite{b,t} and employing the estimates on the wavefunction we showed to hold at horizon crossing, one is able to derive the inequalities below:
\bea
\frac{\bar M_6^2 H^2}{M_2^4(\alpha_0+\sqrt{\beta_0})}>1\,;\qquad \frac{\bar M_6^2 H^2}{M_3^4(\alpha_0+\sqrt{\beta_0})^2}>1\,.   \label{ineq2}
\eea
Again, it is sufficient that only one inequality in Eq.~(\ref{ineq1}) and the corresponding one in Eq.~(\ref{ineq2}) are satisfied. The coefficient $M_4$ is not found in the equation above as it first appears in theory Lagrangian at fourth order, so it is not involved in the tree level bispectrum calculations. The further  freedom on $M_4$ (and other coeffcients) that results from this simple fact can be used to study models of inflation which present a relatively small bispectrum together with a larger trispectrum signal \cite{t,4pt}. The inequalities given in  Eq.~(\ref{ineq1}) and (\ref{ineq2}) above are indeed compatible for values of $(\alpha_0+\sqrt{\beta_0})$ smaller than unity,  an assumption which is generally made when looking for models that can produce large NG, as shown in \cite{b,t}. We see then that there is a whole, large window of values for the coefficient $\bar M_6$ that would allow for a running of $f_{NL}$ dominated by $\epsilon_6$. By looking at Eq.~(\ref{f6}) one realizes the running can indeed be large since it was shown in \cite{b} that $f_{NL}^{\bar M_6}$ itself can give the leading contribution to the total bispectrum amplitude.  This effect is generated purely by third order terms and specifically by extrinsic curvature-generated interaction terms driven by coefficients which first appear in the Lagrangian at third order. Adding the results of this section to the analysis of \cite{b,t} one can safely say that, with respect to all observables one is ultimately interested in, the analysis of these curvature terms has shown they can have leading effects on all quantities and must therefore always be included in a thorough analysis of non-Gaussianities.

\section{Conclusions}
In this work we employed the tools of effective field theory to analyze a very general theory of inflation driven by a single scalar degree of freedom. We have done so at the level of the quadratic and higher-order Lagrangian. The solution to the equation of motion for the scalar has been analyzed in detail. The freezing of the $k$-modes outside the effective horizon is shown to be preserved in this setup as well. In Section 2 we gave the resulting power spectrum at leading order as a function of all possible parameters. We learned that, in order to consider a fully general theory, one has to raise the number of generalized slow-roll parameters to five, as opposed to the usual three ($H$, $c_s$, $\epsilon$).  Naturally, this fact allows for more freedom and we stress here that the two additional degrees of freedom ($\epsilon_0$ and $\epsilon_1$, Eq.~(\ref{genep})) are generated by extrinsic curvature terms in the quadratic Lagrangian that can be significant in specific cases such as near de Sitter limit, or for small values of the generalized speed of sound or a combination thereof. From the spectrum, we proceeded to obtained its tilt and running. In \textit{Sec. 2.3} we also discussed the (mild) bounds on the generalized slow roll parameters. In \textit{Sec. 3} we described a next-to-leading order calculation for the tilt and running of the power spectrum, specialized to the Ghost Inflation case. At third order, we studied the running of the bispectrum amplitude $f_{NL}$ contributed by a curvature-generated interaction term driven by $\bar{M}_6$. From previous investigations \cite{b,t}, it was known that this term is able to generate important or even leading non-Gaussianities and presents distinctive features in the form of the bispectrum and trispectrum shape-function. In the last section we showed that, upon imposing suitable and mild bounds on the coefficients driving the interactions, the running of $f_{NL}$ is dominated by this term without spoiling any of the aforementioned interesting features on the three and four-point function and without affecting the leading value of the power spectrum. In particular, we showed that the running $n_{NG}$ can be such that $n_{NG}\gg {\cal O}(\epsilon,\eta)$.\\ All the above results clearly point to the importance of considering extrinsic curvature generated terms in the quest for predictions on important observable quantities.

\section*{Acknowledgments}
This research has been partially supported by the ASI/INAF Agreement I/072/09/0 for the Planck LFI Activity
of Phase E2. MF would like to thank Claudio Destri for insightful discussions and for constant and
kind encouragement during the completion of this work. MF thanks the
Physics Department of the University of Padova for warm hospitality.

\section*{Appendix}
\subsection*{Explicit expression for the slow-roll parameter $\epsilon_{\Gamma}$}
We give here an explicit expression for the time dependence of the slow-roll parameters which we called $\epsilon_{\Gamma}$. This quantity if first written in terms of the $M, \bar M$ parameters:
\bea
\fl \Gamma[t]= \gamma \left[\frac{5}{4}+\frac{H^2 M_P^2 \epsilon +\frac{1}{2} M_1{}^4}{\left(H^2 M_P^2 \epsilon +2 M_2{}^4+3 M_1{}^4\right) \left(\frac{4 \left(H^2 M_P^2 \epsilon +\frac{1}{2} M_1{}^4\right)}{H^2 M_P^2 \epsilon +2 M_2{}^4+3 M_1{}^4}-4 i \sqrt{\frac{M_0{}^4}{2 \left(H^2 M_P^2 \epsilon +2 M_2{}^4+3 M_1{}^4\right)}}\right)}\right]\times\nonumber\\
\fl \gamma\left[\frac{5}{4}+\frac{H^2 M_P^2 \epsilon +\frac{1}{2} M_1{}^4}{\left(H^2 M_P^2 \epsilon +2 M_2{}^4+3 M_1{}^4\right) \left(\frac{4 \left(H^2 M_P^2 \epsilon +\frac{1}{2} M_1{}^4\right)}{H^2 M_P^2 \epsilon +2 M_2{}^4+3 M_1{}^4}+4 i \sqrt{\frac{M_0{}^4}{2 \left(H^2 M_P^2 \epsilon +2 M_2{}^4+3 M_1{}^4\right)}}\right)}\right],
\eea
where $\gamma[t]$ is the Euler function. From here one calculates the quantity $\epsilon_{\Gamma}=-\frac{\dot \Gamma}{H \Gamma }$, obtaining:
\bea
\fl \frac{\dot \Gamma}{H \Gamma}= \left(i \left(Poly\Gamma\left[0,\frac{5}{4}+\left(4-\frac{8 i \left(H^2 M_P^2 \epsilon +2 M_2{}^4+3 M_1{}^4\right) \sqrt{\frac{M_0{}^4}{2 H^2 M_P^2 \epsilon +4 M_2{}^4+6 M_1{}^4}}}{2 H^2 M_P^2 \epsilon +M_1{}^4}\right)^{-1}\right]\times \right. \right.\nonumber \\
\fl \left. \left.  \left(i \left(2 H^2 M_P^2 \epsilon +M_1{}^4\right)-2 \left(H^2 M_P^2 \epsilon +2 M_2{}^4+3 M_1{}^4\right) \sqrt{\frac{M_0{}^4}{2 H^2 M_P^2 \epsilon +4 M_2{}^4+6 M_1{}^4}}\right){}^2 \right. \right. \nonumber\\
\fl -Poly\Gamma\left[0,\frac{5}{4}+\left(4+\frac{8 i \left(H^2 M_P^2 \epsilon +2 M_2{}^4+3 M_1{}^4\right) \sqrt{\frac{M_0{}^4}{2 H^2 M_P^2 \epsilon +4 M_2{}^4+6 M_1{}^4}}}{2 H^2 M_P^2 \epsilon +M_1{}^4}\right)^{-1}\right]\times \nonumber\\
\fl \left. \left(i \left(2 H^2 M_P^2 \epsilon +M_1{}^4\right)+2 \left(H^2 M_P^2 \epsilon +2 M_2{}^4+3 M_1{}^4\right) \sqrt{\frac{M_0{}^4}{2 H^2 M_P^2 \epsilon +4 M_2{}^4+6 M_1{}^4}}\right){}^2\right)\times \nonumber\\
\fl \left(2 H^4 M_P^4 \epsilon ^2 \frac{d}{dt}\left(\frac{M_0{}^4}{2 H^2 M_P^2 \epsilon +4 M_2{}^4+6 M_1{}^4}\right)+4 H^2 M_P^2 \epsilon  M_2{}^4  \frac{d}{dt}\left(\frac{M_0{}^4}{2 H^2 M_P^2 \epsilon +4 M_2{}^4+6 M_1{}^4}\right) \right.\nonumber \\
\fl +7 H^2 M_P^2 \epsilon  M_1{}^4 \frac{d}{dt}\left(\frac{M_0{}^4}{2 H^2 M_P^2 \epsilon +4 M_2{}^4+6 M_1{}^4}\right) +2 M_2{}^4 M_1{}^4 \frac{d}{dt}\left(\frac{M_0{}^4}{2 H^2 M_P^2 \epsilon +4 M_2{}^4+6 M_1{}^4}\right)\nonumber\\ 
\fl +3 M_1{}^8 \frac{d}{dt} \left(\frac{M_0{}^4}{2 H^2 M_P^2 \epsilon +4 M_2{}^4+6 M_1{}^4}\right)-16 H M_P^2 \epsilon  M_2{}^4 \dot H \frac{M_0{}^4}{2 H^2 M_P^2 \epsilon +4 M_2{}^4+6 M_1{}^4}\nonumber \\
\fl -20 H M_P^2 \epsilon  M_1{}^4 \dot H \frac{M_0{}^4}{2 H^2 M_P^2 \epsilon +4 M_2{}^4+6 M_1{}^4}-8 H^2 M_P^2 M_2{}^4 \dot  \epsilon  \frac{M_0{}^4}{2 H^2 M_P^2 \epsilon +4 M_2{}^4+6 M_1{}^4}   \nonumber \\
\fl -10 H^2 M_P^2 M_1{}^4 \dot \epsilon  \frac{M_0{}^4}{2 H^2 M_P^2 \epsilon +4 M_2{}^4+6 M_1{}^4}+32 H^2 M_P^2 \epsilon  M_2{}^3 \dot M_2 \frac{M_0{}^4}{2 H^2 M_P^2 \epsilon +4 M_2{}^4+6 M_1{}^4} \nonumber \\
\fl +16 M_2{}^3 M_1{}^4 \dot M_2 \frac{M_0{}^4}{2 H^2 M_P^2 \epsilon +4 M_2{}^4+6 M_1{}^4}+40 H^2 M_P^2 \epsilon  M_1{}^3 \dot M_1 \frac{M_0{}^4}{2 H^2 M_P^2 \epsilon +4 M_2{}^4+6 M_1{}^4}      \nonumber \\
\fl -16 M_2{}^4 M_1{}^3 \dot M_1 \frac{M_0{}^4}{2 H^2 M_P^2 \epsilon +4 M_2{}^4+6 M_1{}^4}  +\left(2 H^2 M_P^2 \epsilon +M_1{}^4\right) \left(H^2 M_P^2 \epsilon +2 M_2{}^4+3 M_1{}^4\right)\times \nonumber \\
\fl \left. \left. \frac{M_0{}^3 \left(4 \left(H^2 M_P^2 \epsilon +2 M_2{}^4+3 M_1{}^4\right) \dot M_0-M_0 \left(H M_P^2 \left(2 \epsilon  \dot H+H \dot \epsilon \right)+8 M_2{}^3 \dot M_2+12 M_1{}^3 \dot M_1\right)\right)}{2 \left(H^2 M_P^2 \epsilon +2 M_2{}^4+3 M_1{}^4\right){}^2}\right)\right) / \nonumber\\
\fl \left( H \sqrt{\frac{2 M_0{}^4}{2 H^2 M_P^2 \epsilon +4 M_2{}^4+6 M_1{}^4}} \left(\left(2 H^2 M_P^2 \epsilon +M_1{}^4\right){}^4 +8 \left(2 H^2 M_P^2 \epsilon +M_1{}^4\right){}^2\times \right. \right.\nonumber\\
\fl \left(H^2 M_P^2 \epsilon +2 M_2{}^4+3 M_1{}^4\right){}^2 \frac{M_0{}^4}{2 H^2 M_P^2 \epsilon +4 M_2{}^4+6 M_1{}^4}+16 \left(H^2 M_P^2 \epsilon +2 M_2{}^4+3 M_1{}^4\right){}^4 \times \nonumber \\
\fl \left. \left. \frac{M_0{}^4}{2 H^2 M_P^2 \epsilon +4 M_2{}^4+6 M_1{}^4}{}^2\right)\right).
\eea

\subsection*{Explicit expression for the $\Theta$ functions of Eq.~(\ref{run}).}

\bea
\fl \Theta= \frac{7 H^2 M_P^2 \epsilon+8 \left(2 M_2{}^4+3 M_1{}^4\right) }{2  \left(H^2 M_P^2 \epsilon+2 M_2{}^4+3 M_1{}^4\right)}+\nonumber\\ 
\fl  \frac{8 H^4 M_P^4 \epsilon^2-16 H^2 M_P^2 \epsilon \left(2 M_2{}^4+3 M_1{}^4\right)+H^2 M_P^2 \epsilon  \left(-14 H^2 M_P^2 \epsilon +8 M_2{}^4+15 M_1{}^4\right) }{2  \left(H^2 M_P^2 \epsilon+2 M_2{}^4+3 M_1{}^4\right) \left(2 H^2 M_P^2 \epsilon+M_1{}^4+M_0{}^2 \sqrt{2 H^2 M_P^2 \epsilon +4 M_2{}^4+6 M_1{}^4}\right)};\nonumber\\
\fl \Theta_{\eta}=  \frac{ H^2 M_P^2  \epsilon }{4 \left(H^2 M_P^2 \epsilon +2 M_2{}^4+3 M_1{}^4\right)}+\nonumber\\
\fl \frac{ \left(6 H^2 M_P^2 \epsilon +24 M_2{}^4+33 M_1{}^4\right)H^2 M_P^2 \epsilon }{4 \left(H^2 M_P^2 \epsilon +2 M_2{}^4+3 M_1{}^4\right) \left(2 H^2 M_P^2 \epsilon +M_1{}^4+M_0{}^2 \sqrt{2 H^2 M_P^2 \epsilon +4 M_2{}^4+6 M_1{}^4}\right)};\nonumber\\
\fl\Theta_{2}= \frac{ 2 M_2{}^4}{ \left(H^2 M_P^2 \epsilon +2 M_2{}^4+3 M_1{}^4\right)}+\nonumber\\
\fl- \frac{ \left(6 H^2 M_P^2 \epsilon +3 M_1{}^4\right)2 M_2{}^4 }{ \left(H^2 M_P^2 \epsilon +2 M_2{}^4+3 M_1{}^4\right) \left(2 H^2 M_P^2 \epsilon +M_1{}^4+M_0{}^2 \sqrt{2 H^2 M_P^2 \epsilon +4 M_2{}^4+6 M_1{}^4}\right)};\nonumber \\
\fl \Theta_{1}=  \frac{3 M_1{}^4 }{ \left(H^2 M_P^2 \epsilon +2 M_2{}^4+3 M_1{}^4\right) }+\nonumber\\
\fl \frac{\left(+ 3M_1{}^4-4 H^2 M_P^2 \epsilon +4 M_2{}^4\right)3 M_1{}^4 }{ \left(H^2 M_P^2 \epsilon +2 M_2{}^4+3 M_1{}^4\right) \left(2 H^2 M_P^2 \epsilon +M_1{}^4+M_0{}^2 \sqrt{2 H^2 M_P^2 \epsilon +4 M_2{}^4+6 M_1{}^4}\right)}; \nonumber \\ 
\fl \Theta_{0}=  \frac{6 M_0^2 }{ \left(2 M_0{}^2+\sqrt{2} \left(2 H^2 M_P^2 \epsilon +M_1{}^4\right) \sqrt{\frac{1}{H^2 M_P^2 \epsilon +2 M_2{}^4+3 M_1{}^4}}\right)}. 
\eea
\subsection*{Explicit expression for variables  in Eq.~(\ref{eoma}),(\ref{eomb}).}
\bea
\fl \frac{f^{''}}{f} = 2 a^2 H^2 \left(1-\frac{\epsilon}{2} +\frac{6 M_2{}^3 \dot M_2}{H \left(2 M_2{}^4+3 M_1{}^4-M_P^2 \dot H\right)}+\frac{9 M_1{}^3 \dot M_1}{H \left(-2 M_2{}^4-3 M_1{}^4+M_P^2 \dot H\right)} \right.\nonumber\\
\fl \left.\qquad \qquad \qquad +\frac{3 M_P^2 \ddot H}{4 H \left(-2 M_2{}^4-3 M_1{}^4+M_P^2 \dot H\right)}       \right) \, .
\eea
\bea
\tilde \alpha_0 =\alpha_0(1+\frac{\dot \alpha_0}{H \alpha_0}) \,;\qquad  \tilde \beta_0=\beta_0(1+\frac{\dot \beta_0}{H \beta_0}) \,.
\eea
In going from  Eq.~(\ref{eoma}) to Eq.~(\ref{eomb}) it was safely assumed that $a(\tau)\simeq -\frac{1}{H \tau(1-\epsilon)}$, which in turn means that the cmplete and most general expression for the parameter $x_0$ is, to first order in generalized slow-roll parameters, given by:
\bea
\fl x_0 = \left(-\frac{\epsilon}{2} +\frac{6 M_2{}^3 \dot M_2}{H \left(2 M_2{}^4-3 M_1{}^4-M_P^2 \dot H\right)}+\frac{9 M_1{}^3 \dot M_1}{H \left(-2 M_2{}^4+3 M_1{}^4+M_P^2 \dot H\right)} \right.\nonumber\\
\fl \left.\qquad \qquad \qquad +\frac{3 M_P^2 \ddot H}{4 H \left(-2 M_2{}^4+3 M_1{}^4+M_P^2 \dot H\right)} +2\epsilon  +\frac{3}{2}\epsilon \frac{ M_P^2 {\dot H(t)}}{2M_2^4-M_P^2 \dot H +3 M_1^4 }   \right)\, .
\eea

\vspace{10mm}
\section*{References}
\bibliographystyle{JHEP}

\end{document}